\documentclass[12pt]{article}
\usepackage{graphicx}
\usepackage{amssymb}
\usepackage{amsmath}
\usepackage{xcolor}

\def\al{\alpha}
\def\be{\beta}
\def\ga{\gamma}

\def\de{\delta}

\def\eps{\epsilon}

\def\la{\lambda}
\def\si{\sigma}

\def\om{\omega}

\def\La{\Lambda}
\def\la{\lambda}
\def\vp{\varphi}
\def\ve{\varepsilon}
\def\pa{\partial}
\def\ra{\Rightarrow}

\def\bn{{\bf{n}}}
\def\bee{{\bf{e}}}
\def\bmu{{\boldsymbol{\mu}}}

\def\two{{\dot 2}}

\def\bg{{\bar{g}}}

\def\cA{{\mathcal A}}

\def\cD{{\mathcal D}}
\def\cF{{\mathcal F}}

\def\cM{{\mathcal M}}

\def\cR{{\mathcal R}}

\def\re{{\rm e}}

\def\rd{{\rm d}}

\def\rR{{\cR}}

\def\rT{{\rm T}}

\def\beq{\begin{equation}}
\def\eeq{\end{equation}}
\def\bea{\begin{eqnarray}}
\def\eea{\end{eqnarray}}
\def\nn{\nonumber}

\def\ra{\rightarrow}
\def\Ra{\Rightarrow}

\begin{document}
\begin{center}
{\bf\Large Fundamental Membranes \\[5mm] and the String Dilaton}\\[9mm]
{\bf Krzysztof A. Meissner$^1$ and Hermann Nicolai$^2$}\\[8mm]
{$^1$Faculty of Physics,
University of Warsaw\\
Pasteura 5, 02-093 Warsaw, Poland\\
$^2$Max-Planck-Institut f\"ur Gravitationsphysik\\
(Albert-Einstein-Institut)\\
M\"uhlenberg 1, D-14476 Potsdam, Germany\\
}
\end{center}

\vspace{1cm}

\begin{abstract} 
\noindent
We study the quantization of the bosonic sector of supermembrane theory in 
double dimensional reduction, in order to extract the dependence of the resulting  
world-sheet action on the string dilaton (which cannot be obtained from a purely 
kinematic reduction). Our construction relies on a Polyakov-type approach with 
all six metric components on the world-volume as independent quantum 
fields, and shows that the correct and unique answer is only obtained 
if the target-space dimension of the theory is restricted  to the critical 
value ($D=11$ for the supermembrane). As a corollary, our analysis implies that 
there are no analogs of the non-critical string for (super-)membrane theory. 
\end{abstract}

\vspace{5mm}
\noindent

\newpage
\section{Introduction}

Membranes are notoriously hard to quantize. This is because they are defined by 
world-volume actions that, unlike the world-sheet actions of (super-)string theory, 
involve seemingly intractable interactions on the world-volume. For this reason it
is not even  clear whether and under what circumstances a quantum (super-)membrane 
theory can be sensibly defined at all. Nevertheless, and intriguingly, the maximally 
extended $D=11$ supermembrane theory \cite{BST,BST1} stands out uniquely 
as a challenging candidate for the {\em non-perturbative} quantum unification of 
gravity and  matter, with maximally extended $D=11$ 
supergravity \cite{CJS} as a `low energy limit'.

Early work on the quantization of the bosonic membrane in a Minkowski
target-space background relied on a Hamiltonian formulation in the light-cone 
gauge and demonstrated its equivalence with the $N\ra\infty$ limit of a certain 
SU($N$) matrix model \cite{G,Hoppe}. Building on these insights it was 
shown in \cite{dWHN} that the light-cone gauge formulation of the supermembrane 
(again in flat target-space) leads to a maximally supersymmetric matrix model,
corresponding to the reduction of a maximally supersymmetric Yang-Mills
theory to one (time) dimension. This result greatly improved prospects for 
making the supermembrane amenable to a quantum treatment which can 
accommodate $D=11$ supergravity as a massless sector. Indeed, nine years 
later the very same model was proposed as a model of M theory \cite{BFSS}.
A crucial role here is played by the fact that the spectrum of the 
supersymmetric Hamiltonian is continuous \cite{dWLN,Smilga}. Altogether, 
these developments clarified that the supermembrane,  unlike (super-)string 
theory, does not admit a proper first quantized formulation, but must be 
regarded as a second quantized theory from the outset \cite{HN}.
For a review of these developments and for different perspectives
on them, see \cite{HN,Duff,deWit,Banks,Taylor}.

Nevertheless, despite these advances there has overall been only scant progress 
on the quantization of the supermembrane and the $N\ra\infty$ limit of the matrix 
model, with major results mainly concerning the spectrum of the 
Hamiltonian and the existence (or non-existence) of a normalizable ground 
state wave function for fixed finite $N$ (see \cite{FH,MNS,R1,Hoppe1,R2} 
and references therein),  and the construction 
of (classical analogs of) superstring vertex operators for the massless 
supermembrane states \cite{DNP}. 
Furthermore, very recent work \cite{LN} has provided 
evidence that supersymmetry is a necessary prerequisite for the 
$N\ra\infty$ limit of the matrix model to  exist, thus lending credence to old claims
that the bosonic membrane is not renormalizable in any dimension. However,
all this work pertains to the quantization of membrane theory in a {\em flat} 
target-space background only, and does not generalize in any obvious way 
to curved target-space geometries.

In this paper we return to the Lagrangian formulation of the membrane
and supermembrane in an arbitrary curved target-space background,
focusing on the links between membrane theory and the 
string dilaton. As is well known, the latter occupies a central place in
string theory via its direct relation to the string coupling $g_s$.
In a first step we here study the double dimensional reduction of 
the supermembrane and its quantization in a Polyakov-type
formulation, with the world-volume metric as independent quantum variables.
As shown in \cite{Duff1,AKS} the kinematical reduction together with the 
embedding equations reproduces part of the type IIA superstring 
action in a Green-Schwarz formulation, in particular with the correct 
world-sheet couplings of the target-space metric $G^{(10)}_{\mu\nu}(X)$ 
and the two-form field $B_{\mu\nu}(X)$. While
the membrane theory has only one parameter, the membrane tension $\rT_3$,
there appears a second dimensionful parameter in the double dimensional 
reduction, the circumference $R_{10}$ of the compactified direction. 
We then have the relations
\beq\label{TsT3}
\rT_s  \equiv (2\pi \alpha')^{-1}\,=\,  R_{10} \rT_3  \quad ,\quad g_s^2 \,=\, R_{10}^3 \rT_3
\eeq
where $\rT_s$ is the string tension and $g_s$ is the
string coupling \cite{T,Russo}.
For fixed $\rT_3$ these relations in particular imply the well 
known result $R_{10} \propto g_s^{2/3}$ \cite{Witten}.
However, apart from the (anticipated) disappearance of the Ramond-Ramond 
fields in this reduction, there remains no trace of the remaining missing 
piece of the type I subsector in the resulting Lagrangian, namely the dilaton!
More specifically, the origin of the crucial term \cite{FT}
\beq
\frac{1}{4\pi} \int \rd^2\si \, \sqrt{g}\,  \phi\,  \rR (g)
\label{phiR}
\eeq
(where $\rR\equiv \rR^{(2)}$ is the world-sheet curvature) 
giving rise to the identification between the string coupling $g_s$ 
and the dilaton vacuum expectation value remains unexplained.
We note, however, that for constant $\phi$ a derivation
of this term was already proposed in \cite{BP}.

The main purpose of this work is to argue that the derivation 
of (\ref{phiR}) and a proper understanding of the dilaton couplings on 
the world-sheet require a {\em quantum treatment} of the membrane, 
beyond the classical considerations of \cite{Duff1}, and an approach where one 
integrates over {\em all} six world-volume degrees of freedom
(see also \cite{U} for an alternative approach). In addition
there is the peculiar feature that the field $\phi$ in (\ref{phiR}) requires
a special wave-function renormalization from the membrane perspective,
as we will explain in section~4. These conclusions are
consistent with the fact that the term (\ref{phiR}) does not come
with a factor $(4\pi \al')^{-1}$, unlike the tree level couplings. Our construction 
furthermore reveals the necessity of restricting the target-space 
dimension to  a critical value as a consistency condition (which, however,
may not be sufficient), leading to the conclusion that the membrane and the 
supermembrane can be viable, if they are viable at all, only in target-space 
dimensions $D=27$ and $D=11$ (see also \cite{MKS} for very early work 
where the same conclusion was reached by different arguments). 


Putting together the available evidence we thus conclude that, if at all, only the 
{\em maximally extended} supermembrane theory can give rise to a viable quantum 
theory, because (1) $D=11$ is a necessary condition by the results of the present
work, and (2) supersymmetry is required by the arguments of \cite{LN}.
This implies that, unlike for string theory, there appears to be no such thing 
as `non-critical (super-)membrane theory'!

The structure of this paper, then, is as follows. In section 2 we review the kinematics of
double dimensional reduction, following \cite{Duff1,AKS}. In section~3 we analyze the 
quantization of the theory in a Polyakov-type approach. Finally in section 4 we show
that the usual dilaton coupling is the only sensible, and in fact unique, outcome. 
Although our main concern is the supermembrane we restrict attention
to its bosonic subsector, as the fermionic terms do not affect our main conclusions.

\section{Double dimensional reduction}

We start from the bosonic world-volume action in `Polyakov form' \cite{BdVH,DZ}  
with Euclidean signature 
\bea\label{S}
S \,&=&\, \frac{\rT_3}2\int\rd^3\si\sqrt{\ga}
\Big(\ga^{ij}\pa_i X^M \pa_j X^N G_{MN}(X) -1\Big)  + \nn\\[2mm]
&& \qquad + \,\frac{\rT_3}6 \int \rd^3 \si \, \ve^{ijk} \pa_i X^M \pa_j X^N \pa_k 
X^P \cA_{MNP}(X)
\eea
where $i,j=0,1,2$ and $M,N=0,\ldots,10$ with the world-volume coordinates
$\si^i \equiv (\si^0, \si^1, \si^2)$; below we will occasionally write $\xi \equiv \si^2$ to 
distinguish the compactified coordinate. $G_{MN}(X)$ is the (in general curved) target 
space metric;  $\cA_{MNP}(X)$ is the 3-form field of $D=11$ supergravity. 
For flat target (Minkowski) space we have $G_{MN} = \eta_{MN}$.
The only dimensionful parameter is the membrane tension $\rT_3$, of mass 
dimension three. An alternative way of writing $S$ in terms of flat target-space indices is
\bea\label{Sflat}
S \,&=&\, \frac{\rT_3}2\int\rd^3\si\sqrt{\ga}
\Big(\ga^{ij} \Pi_i^{\;A}  \Pi_j^{\;B} \eta_{AB} -1\Big)  + \nn\\[2mm]
&& \qquad\qquad +  \,\frac{\rT_3}6 \int \rd^3 \si \, \ve^{ijk} \Pi_i^{\;A}  \Pi_j^{\;B}
\Pi_k^{\;C} \cA_{ABC}
\eea
where
\beq\label{Pi}
\Pi_i^{\;A} \,\equiv \, \pa_i X^M E_M^{\;\;A}(X)      
\eeq
with the target-space elfbein $E_M^{\;\;A}$ and flat (Lorentz) indices $A,B,C,...$.
Both actions (\ref{S}) and (\ref{Sflat}) are manifestly invariant under
world-volume diffeomorphisms. Because $\pa_i X^M$ transforms as a vector 
in target-space, they are likewise invariant under target-space diffeomorphisms.
Finally, the action is invariant under the target space gauge
transformations $\de\cA_{MNP} = 3 \pa_{[M} \La_{NP]}$.

The solution of the (algebraic) equations of motion for $\ga_{ij}$ is
\beq\label{gamma}
\ga_{ij}=\pa_i X^M \pa_j X^N G_{MN}(X)
\eeq
so $\ga_{ij}$ is just the induced metric on the world-volume; note that the
second term in (\ref{S}) does not depend on the world-volume metric, hence 
does not contribute to the embedding equations of motion.  Varying the 
target-space coordinates gives
\bea\label{EoMX}
&& \pa_i \Big(\sqrt{\ga}\ga^{ij} \pa_j X^M \Big) \,+\,  
\sqrt{\ga} \ga^{ij} \Gamma^M_{\;PQ}(X)  \pa_i X^P \pa_j X^Q  \; +  \nonumber
 \\[2mm]
&& \hspace{1.5cm} + \;\frac16 \, \ve^{ijk} \pa_i X^N \pa_j X^P \pa_k X^Q \cF^M{}_{NPQ}(X) \,=\, 0
\eea
where $\cF_{MNPQ} \equiv 4 \pa_{[M} \cA_{NPQ]}$, and $\Gamma^M_{\; PQ}$ is
the affine connection associated to the target space metric $G_{MN}(X)$.
There is no other equation  that relates $\cA_{MNP}$ to world-volume objects. 
As explained in \cite{BST,BST1} the supersymmetry of the supermembrane 
action requires that both $G_{MN}$ and $\cA_{MNP}$ satisfy their respective
target superspace equations of motion. 

Even before dimensionally reducing the theory, the world-volume metric can 
be decomposed via the standard Kaluza-Klein (KK)  ansatz
\bea
\ga_{ij}  \,&=& \, \begin{pmatrix}
g_{\al\be} +\re^{2\tau} A_\al A_\be \,&\, \re^{2\tau}A_\al\\
\re^{2\tau}A_\be \, &\,  \re^{2\tau}
\end{pmatrix},        \nn\\[2mm]
\ga^{ij} \,&=&\,
\begin{pmatrix}
g^{\al\be} \,&\, - A^\al\\
- A^\be \,&\,  \re^{-2\tau}+ A^\ga A_\ga
\end{pmatrix},\ \ \ 
\label{gachoice}
\eea
where the world-volume indices are split as $i = (\al,\two)$ {\em etc.}
into world-sheet indices $\al,\be,...$ and the remaining third coordinate
(we sometimes put a dot on the last index to indicate that it is a 
curved index). $g_{\al\be}$ is the $2\times 2$ world-sheet metric, $A_\alpha$ 
is a 2-vector and  $\rho \equiv \re^\tau$ is the world-volume dilaton. Then
\beq
\sqrt{\ga} = \re^\tau\sqrt{g}
\eeq

The double dimensional reduction scheme, or `DDR' for short, in part 
reproduces the string or 
superstring (Green-Schwarz) worldsheet actions by a simple kinematic reduction
that makes partial use of the embedding equations (\ref{gamma}), and
is implemented by identifying the 10th target-space coordinate 
$X^{10}$ with the third world-volume coordinate $\si^2\equiv \xi$ \cite{Duff1}.
Here we slightly generalize this ansatz by assuming the membrane world-volume
to be of the following topological shape (see also \cite{BP})
\beq
\mbox{world-volume} \quad \sim \quad \Sigma_n \times S^1
\eeq
where $\Sigma_n$ is a Riemann surface of genus $n$.  Then we set
\beq
X^\mu = X^\mu(\si^0,\si^1) \quad\mbox{for $\mu =0,...,9\;\;$}  
\label{Xchoice}
\eeq
and identify the 10th target-space coordinate with the third world-volume coordinate
according to~\footnote{In principle we could also use
\beq
 \pa_\al X^{10} = v_\al    \nn
\eeq
with $v_\al$ a harmonic vector field on $\Sigma_n$, which obeys
$\pa_\al v_\be - \pa_\be v_\al = 0$ and $\pa_\al(\sqrt{g} g^{\al\be} v_\be) = 0$;
there are $2n$ independent such vector fields on $\Sigma_n$ 
This means that in principle, the target space coordinate $X^{10}$ can wrap 
around not only $S^1$, but simultaneously around any non-trivial cycle on the 
Riemann surface.}  
\beq
\pa_\two X^{10} = 1\;\;,\quad \pa_\al X^{10} \,=\, 0 
\eeq

For the target-space metric we proceed again from the standard KK ansatz 
\beq\label{11metric}
G^{(11)}_{MN}  \,=\,
\begin{pmatrix}
\re^{-\frac23 \phi} G^{(10)}_{\mu\nu}  + \re^{\frac43 \phi} \cA_\mu \cA_\nu & 
\; \re^{\frac43 \phi} \cA_\mu \\[2mm]
\re^{\frac43 \phi} \cA_\nu & \; \re^{\frac43 \phi}
\end{pmatrix}
\eeq
where $\mu,\nu=0,\ldots,9$, and $ G^{(10)}_{\mu\nu} (X) 
= e_\mu^{\:\;a}(X) e_{\nu a}(X)$ and $\cA_\mu \equiv \cA_\mu(X)$.  
The associated elfbein (in triangular gauge) is
\beq\label{11bein}
E_M^{\; \;A}  \,=\,
\begin{pmatrix}
\re^{-\frac13 \phi} e_\mu^{\;\;a}   & 
\; \re^{\frac23 \phi} \cA_\mu \\[2mm]
0  & \; \re^{\frac23 \phi}
\end{pmatrix} \quad \Rightarrow \quad
E_A^{\; \;M}  \,=\,
\begin{pmatrix}
\re^{\frac13 \phi} e_a^{\;\; \mu}   & 
\; - \, \re^{\frac13 \phi} \cA_a \\[2mm]
0  & \; \re^{- \frac23 \phi}
\end{pmatrix}
\eeq
From these relations and the definition (\ref{Pi}) we read off that
\bea\label{Pi1}
\Pi_\al^{\;a} \,&=&\, \re^{-\frac13 \phi} \pa_\al X^\mu e_\mu^{\;\; a} \quad ,
\quad \Pi_\two^{\; a} \,=\, 0    \nn\\[2mm]
\Pi_\al^{\; 10} \,&=& \, \re^{\frac23 \phi}  \pa_\al X^\mu \cA_\mu     
\quad , \quad \Pi_\two^{\; 10} \,=\, \re^{\frac23 \phi}
\eea
The dilatonic prefactors in (\ref{11metric}) and (\ref{11bein}) have been adjusted 
in order to end up with the standard bosonic action of $D=10$ type IIA supergravity 
in string frame after dimensional reduction from $D=11$ to $D=10$ \cite{Witten} 
\bea
S_{11} \,&=& \, \int \rd^{11} X \, \sqrt{-G^{(11)}} \Big( \cR^{(11)} - \frac1{48} \cF^{MNPQ} \cF_{MNPQ} \Big)
\quad \ra  \\[2mm]
\,&\ra&\,  \int \rd^{10}X \sqrt{- G^{(10)}} \, \re^{-2\phi} \Big( \cR^{(10)} 
- \frac1{12} H^{\mu\nu\rho} H_{\mu\nu\rho} + 4 \, G^{(10)\mu\nu} \pa_\mu\phi \pa_\nu \phi \Big)
\nn\\[2mm]
&& \quad + \, \int \rd^{10} X \,\sqrt{-G^{(10)}} \Big( - \frac14 \cA^{\mu\nu} \cA_{\mu\nu} 
   - \, \frac1{48} \cF^{\mu\nu\rho\si} \cF_{\mu\nu\rho\si} \Big) \,\equiv\, S_{10}    \nn
\label{dimred}
\eea
With this normalization, the terms involving the Ramond-Ramond fields $\cA_\mu$ 
and $\cA_{\mu\nu\rho}$ carry no dilaton factors in the effective target-space 
Lagrangian (see also \cite{Ts}).

The relation (\ref{gamma}) implies in particular
\beq\label{ga22}
\ga_{\two\two} \,\equiv \,  \re^{2\tau} \,=\, \pa_\two X^M \pa_\two X^N G_{MN}
\,=\, \re^{\frac43 \phi}
\eeq
from which we read off the on-shell relation between the world-volume 
dilaton $\tau$ and the target-space dilaton $\phi$
\beq\label{tau}
\tau (\si) \,=\, \frac23 \phi(X(\si))
\eeq
The embedding formula (\ref{gamma}) furthermore requires
\beq\label{A}
\ga_{\al \two} \,\equiv \, \re^{2\tau} A_\al \,=\,  \pa_\al X^M \pa_\two X^N G_{MN} \quad \Ra \quad
A_\al  = \pa_\al X^\mu \cA_\mu    
\eeq
whence on-shell the world-volume KK vector field $A_\al$ gets identified with 
the the pull-back of the target-space KK vector field $\cA_\mu$.  

Next we substitute (\ref{gachoice}) and (\ref{11metric}) into (\ref{Sflat}); for the first 
term on the r.h.s. this gives
\bea\label{S11flat}
S_{\rm DDR}  &=&  \nn\\[2mm]
&& \hspace{-12mm}
 \frac{\rT_3}2   \int \rd\xi \int \rd^2\si \, \re^\tau \sqrt{g}\, \eta_{AB} 
\Big( \ga^{\al\be} \Pi_\al^{\;A} \Pi_\be^{\; B}  + 2\ga^{\al \two} \Pi_\al^{\; A} \Pi_\two^{\; B}
+ \ga^{\two\two} \Pi_\two^{\;A} \Pi_\two^{\; B}  - 1\Big) \nn \\
\eea
The circumference of the compactified dimension is 
\beq
R_{10} = \int \rd \xi \, 
\eeq
and using the relations (\ref{TsT3}),
(\ref{gachoice}) and (\ref{Pi1}) we arrive  
at~\footnote{An equivalent formula was already derived in \cite{AKS}.}
\bea\label{offshellS}
S_{\rm DDR}  \,&=&\, \frac{1}{4\pi\al'}\int \rd^2\si\, 
\bigg[  \re^{\tau - \frac23 \phi} \sqrt{g} g^{\al\be} \pa_\al X^\mu \pa_\be 
X^\nu G^{(10)}_{\mu\nu}(X)  \nn\\[2mm]
&& \hspace{2cm} + \; \re^{\tau + \frac43 \phi} \sqrt{g} g^{\al\be} 
\Big(\pa_\al X^\mu \cA_\mu - A_\al\Big) 
      \Big(\pa_\be X^\nu \cA_\nu - A_\be\Big) \nn\\[2mm]
&&   \hspace{3cm} + \; 2 \; \re^{\frac23 \phi} \sqrt{g} \sinh\left( \frac23 \phi - \tau\right) \bigg]      
\eea
Now substituting the embedding equations (\ref{tau}) and (\ref{A})
we see that the cosmological constant and the terms with KK vector fields 
cancel, and we are left with the canonical string $\si$-model world-sheet action 
\beq\label{S10}
S = \frac{1}{4\pi\al'}\int \rd^2\si\, \sqrt{g} g^{\al\be} \pa_\al X^\mu \pa_\be X^\nu G^{(10)}_{\mu\nu}(X)
\eeq
The dependence on the dilaton field in (\ref{S10}) thus drops out
on-shell, that is, upon partial use of the embedding equations 
(\ref{gamma}). Similarly, the dilaton decouples from
\bea\label{S10H}
&&\frac{\rT_3}6 \int \rd \xi \int \rd^2 \si \, \ve^{ijk} \pa_i X^M \pa_j X^N \pa_k X^P \cA_{MNP}(X)\nn\\[2mm]
&& \hspace{2cm}
= \; \frac1{4\pi \al'} \int \rd^2 \si\, \ve^{\al\be} \pa_\al X^\mu \pa_\be X^\nu B_{\mu\nu}(X)
\eea
but without the use of any embedding equations, since
both $\ve^{ijk}$ and $\ve^{\al\be}$ are densities;
also, $B_{\mu\nu} \equiv \cA_{\mu\nu 10}$.
There is thus again no dilaton dependence in this term.

We therefore reach the conclusion that in this purely kinematical reduction
of the membrane action (\ref{S}) the dilaton disappears altogether from the 
string world-sheet action, even when restricting to the type I subsector.
Likewise the Ramond-Ramond fields $\cA_\mu$ and $\cA_{\mu\nu\rho}$  
drop out, but this was already clear from the well known fact that their inclusion 
requires an extension of the usual NSR formalism.

\section{Quantization}

This leaves the question how to generate the world-sheet dilaton coupling (\ref{phiR})
from the membrane action (\ref{S}).
As is well known, in string theory this term is added in a somewhat 
{\em ad hoc} fashion in order to ensure conformal invariance (that is, 
vanishing $\be$-functions for the target space fields)  also for non-Ricci
flat target space geometries \cite{FT,Callan}. It is important that from 
the string theory perspective, this term should be interpreted as a $\si$-model 
one-loop term, because it comes without a factor of $(4\pi\al')^{-1}$ \cite{FT,Callan}. 
The argument explains why 
(\ref{phiR}) cannot be derived by a simple kinematical reduction from the membrane 
action, unlike (\ref{S10}) and (\ref{S10H}). Here we wish to argue that 
in order to recover the correct dilaton dependence we must replace 
the above classical (on-shell) treatment by a {\em quantum mechanical
(off-shell) treatment} where instead of using the embedding equations (\ref{tau})
and (\ref{A}) one must keep the world volume fields $\tau$ and $A_\al$ as quantum 
fields with the action (\ref{offshellS}) and integrate over them. Such an approach is
evidently much closer in spirit to a Polyakov-type treatment in the sense that 
all six components of the world-volume metric are to be integrated over. This
is also a main difference with \cite{BP} where a more restricted form is assumed
for the world-volume metric, {\em cf.} their eqn.~(11).

In order to implement the integration over all world-volume degrees of freedom 
in (\ref{offshellS}) we decompose the full functional measure 
of the world-volume theory according to
\bea\label{measure1}
\big[\cD \ga_{ij}\big] \, &=& \, \big[\cD g_{\al\be}\big] \big[\cD A_\al\big
]\big[\cD \re^\tau\big]  \nn\\[2mm]
\big[\cD X\big] \, &=& \, \big[\cD^{(10)} X\big] \big[\cD X^{10}]
\eea
where $[\cD X^{(10)}] \equiv [\cD X^0] \cdots [\cD X^9]$.
One new ingredient originating from the DDR of the membrane theory
that is absent from the string world-sheet theory 
is the extra gauge invariance
\beq\label{Xi}
\de X^{10}\,=\, \Xi(X^0,\dots,X^9)
\eeq
that follows from the invariance of the original membrane action (\ref{S})
under target-space diffeomorphisms: indeed, such fluctuations drop out entirely
from (\ref{offshellS}). Now, as is well known from general KK theory, 
the diffeomorphisms along the compactified direction descend to gauge 
transformations on the KK vector in the compactified theory according to
\beq\label{gaugetrafo0}
\de \cA_\mu(X)  \,=\, \pa_\mu \Xi(X)
\eeq
Likewise the world-volume diffeomorphisms split into world-sheet
diffeomorphisms and the KK gauge transformations
\beq\label{gt}
\de A_\al (\si)\,=\, \pa_\al \om (\si)
\eeq
On-shell, the transformations (\ref{gaugetrafo0}) and (\ref{gt})
are identified by (\ref{A})
\beq\label{gaugetrafo}
\de A_\al \, =\,  \pa_\al \om \,\equiv \, \pa_\al X^\mu \pa_\mu \Xi 
\eeq
where $\om(\si) \equiv \Xi(X(\si))$ is the KK gauge transformation parameter 
induced on the world-volume. This argument also shows in what sense 
(\ref{offshellS})  is invariant under world-volume gauge transformations: any gauge 
transformation on the world-volume KK vector $A_\al$ can be absorbed 
into a target-space gauge transformation, where the function $\Xi(X)$ must
coincide with $\om$ on the world-volume for $X^\mu = X^\mu(\si)$, but is 
otherwise arbitrary. Because $\Xi$ is thus a gauge 
transformation parameter in the effective target-space theory, it must
not be integrated over. Consequently, we can remove $[\cD X^{10}]$ from 
(\ref{measure1}), and thus replace the last line of (\ref{measure1}) by
\beq
\big[\cD X\big] \, \longrightarrow \, \big[\cD^{(10)} X\big]
\eeq

Next we observe that the first two lines on the r.h.s. of (\ref{offshellS}) depend only 
on the unimodular part $\sqrt{g} g^{\al\be}$ of the  world-sheet metric, and therefore
the conformal factor $\re^\la\equiv\sqrt{g}$ appears only in the last 
line of the action (\ref{offshellS})  as a linear factor. 
However, it is well known from string theory that a hidden 
dependence on the conformal factor may nevertheless arise via the functional 
measure in the form of a Liouville action \cite{Polyakov,Friedan,OA,MN,FMS,DK}. 
This hidden dependence {\em disappears only in the critical dimension}.
More precisely, only in the latter case we can exploit the invariance of 
the full functional measure under conformal rescalings         
\beq\label{rescale1}
\big [\cD g_{\al\be}\big] \big[\cD^{(10)}X\big] \Big|_{g = \hat{g} \re^\la} \,=\,
 \big[\cD \hat{g}_{\al\be}\big] \big[\cD^{(10)} X\big]\Big|_{\hat{g}}  
 \eeq
To divide out world-sheet diffeomorphisms we follow the standard procedure
(which is beautifully explained in the original papers \cite{Friedan,OA,MN}) by 
parametrizing the metric variations as
\beq
\de g_{\al\be} \,=\, \de\la \, g_{\al\be} \,+\, (Pv)_{\al\be} \,+\,
\de\bmu_r \Psi^{(r)}_{\al\be}
\eeq
with the traceless world-sheet diffeomorphisms
\beq
(P v)_{\al\be} \,:=\, \nabla_\al v_\be + \nabla_\be v_\al - 
g_{\al\be} \nabla_\ga v^\ga
\eeq
Here $\bmu$ coordinatize the moduli space of $\Sigma_n$, and the $\Psi^{(r)}_{\al\be}$
form an orthonormal basis of $({\rm Im}\, P)^\perp = {\rm ker}\, (P^\dagger)$. 
Then it is shown \cite{OA,MN} that after dividing out the diffeomorphisms
the measure can be represented in the form
\beq\label{measure2}
\big[  \cD g_{\al\be} \big] \; \ra \;  \big[\cD \re^\la \big]\,\rd\bmu\,
\big({\det}' P_\bg^\dagger P_\bg\big)^{1/2} \, {\det}^{1/2} H(P_\bg^\dagger)
\eeq
where $\rd \bmu$ is a measure on the moduli space $\cM(\Sigma_n)$,
with $\bg_{\al\be} \equiv \bg_{\al\be}(\bmu)$ a representative metric in the 
diffeomorphism and conformal equivalence class of metrics, and  the finite-dimensional 
matrix $H_{rs}(P^\dagger) = \langle \Psi^{(r)}|\Psi^{(s)}\rangle$ \cite{OA,MN}.
The determinant factors in (\ref{measure2}) are finite functions of the modular
parameters $\bmu$ \cite{OA,MN} which we will disregard in the remainder.
Importantly, $[\cD \re^\la]$ is still part of the measure, but crucially 
there is no other hidden dependence on the conformal factor in the full measure 
by (\ref{rescale1}), so we can take the operator $P_\bg$ to depend on any 
representative metric $\bg_{\al\be}(\bmu)$. Without any other dependence on the
conformal factor -- as in critical string theory -- we can then divide the full measure by the 
(infinite) volume of the group of Weyl rescalings, and thus drop the integral 
over $[\cD \re^\la]$ altogether.

Here the situation is different because there still remains an explicit dependence
on the conformal factor in (\ref{offshellS}). The necessity of restricting the target-space 
dimension to a critical value also for the (super-)membrane is now a consequence
of the fact that only in this case the conformal factor can act as a Lagrange
multiplier field and does not acquire any dynamics 
of its own (as in Liouville theory), so that the integration over 
$[\cD \re^\la]$ can be explicitly performed to produce a 
$\de$-functional $\propto \de\big(\tau -\frac23 \phi\big)$. Once this $\de$-functional
is in place we can finally do the integral over $[\cD \re^\tau]$, which identifies
the world-volume and the target-space dilaton fields also for the quantized 
theory, thus ensuring that the world-volume dilaton likewise does not develop 
any independent dynamics of its  own.  Altogether this leaves us with the world-sheet 
action (the coupling (\ref{S10H}) emerges from the kinematical reduction as before)
\bea\label{offshellS1}
S \,&=&\, \frac{1}{4\pi\al'}\int \rd^2\si\, 
\sqrt{\bg} \bg^{\al\be} \bigg[ \pa_\al X^\mu \pa_\be 
X^\nu G^{(10)}_{\mu\nu}(X)  \; +\nn\\[2mm]
&& \hspace{2cm} + \; \re^{2 \phi} \Big(\pa_\al X^\mu \cA_\mu - A_\al\Big) 
      \Big(\pa_\be X^\nu \cA_\nu - A_\be\Big) \bigg]
\eea
which still depends on the target-space dilaton $\phi\,$.
From the membrane perspective, the conformal (Weyl) invariance 
of the world-sheet theory can thus be viewed as the result of integrating 
out the conformal factor, such that we are only left with the dependence
on the representative metric $\bg_{\al\be}(\bmu)$ and a finite-dimensional
integral over the moduli space $\cM(\Sigma_n)$ (see also \cite{AKS,U} for
a somewhat different perspective on the emergence of conformal
symmetry on the world-sheet).

We now recognize the necessity of restricting the target space dimension to  
a critical value also for the (super-)membrane, from a perspective that is
quite different from string theory. Our arguments imply $D=27$ and $D=11$ 
as necessary (but not sufficient) consistency conditions for the bosonic membrane 
and the supermembrane to exist: only with this assumption, the extra world-volume
degrees of freedom remain kinematical (Lagrange multiplier) degrees of 
freedom without dynamics of their own. With any other choice we would 
be left with a {\em de facto} intractable path integral! These arguments 
also imply that there is no such thing as non-critical membrane theory.

\section{Dilaton coupling}

However, we are not yet done since it remains to integrate over $A_\al$, and
thus to determine the dilaton dependence of the final result. For the
normalization of the integral we choose
\beq\label{intA}
\int \big[ \cD A'_\al\big]_\bg \, \re^{ -  |\!| A'|\!|_\bg^2} \,=\, 1
\eeq
with
\beq
|\!| A' |\!|_\bg^2 \,\equiv\, \frac1{4\pi\al'}
\int \rd^2\si \, \sqrt{\bg} \bg^{\al\be}  A'_\al\,  A'_\be
\eeq
and the redefined (gauge invariant) field $A_\al' \equiv A_\al - \pa_\al X^\mu \cA_\mu$. 
Our aim is to compute the (partial) renormalized effective action functional  
$W = W[\bg_{\al\be},\phi]$
\beq\label{W}
\re^{- W[\bg,\phi]} \,=\, 
\int \big[ \cD A_\al\big]_\bg \, \exp \left( - \frac1{4\pi\al'}\int\rd^2\si \, \re^{2\phi}
\sqrt{\bg} \bg^{\al\be} A_\al A_\be \right)
\eeq
where we drop primes from now on. It is not immediately obvious how to get 
a well-defined answer from this expression, but we now demonstrate, 
subject to the somewhat unusual  renormalization prescription 
(\ref{Ren}) below, that if there  is any sensible answer at all, it must be 
proportional to (\ref{phiR})! 

The key observation is that (\ref{W}) is an ultralocal Gaussian integral which 
can be done explicitly, apart from questions related to the continuum limit.
In order to analyze it, let us discretize the r.h.s. of (\ref{W}), with two-dimensional 
lattice points $a\bn$ on (a local coordinate patch of) the discretized Riemann surface, 
and lattice spacing $a$ and $\bn\in\mathbb{Z}^2$. We wish to calculate the $a\ra 0$ 
limit of the integral
\bea
\re^{-f_a(\phi)} \,&=&\, \int \prod_{\bn\,;\,\al=0,1}  \left[ \frac{a\cdot dA_\al(a\bn)}{2\pi\sqrt{\al'}} \right] \; \times  \nn\\[2mm]
&& \hspace{-1cm} \times \; \exp\left( - \frac{a^2}{4\pi\al'} 
\sum_\bn \re^{2\phi(a\bn)} G^{\al\be}(a\bn)  A_\al(a\bn) A_\be (a\bn)\right)
\eea
where $G^{\al\be}(a\bn) \equiv  \sqrt{\bg}  \bg^{\al\be}(a\bn)$ is unimodular;
because of unimodularity no further normalization is required, and the
integral is thus normalized in such a way that $f_a(0)= 0$. 
Now by rescaling integration variables it is easy to see that for
non-vanishing $\phi(a\bn)$ we have, on the given patch,
\beq\label{f}
f_a(\phi) \,=\, 2 \sum_\bn \phi(a\bn)
\eeq
As desired, this is linear in $\phi$,  so the only question is how to
perform the continuum limit $a\ra 0$ in such a way as to get a sensible
and well-defined result. However, as it stands, the limit $a\ra 0$ does {\em not} 
exist because we are lacking a prefactor $a^2$ for this expression to be converted
into a Riemann sum for an integral. The only way to remedy this situation
is to insert a factor $a^2 \times (Z(a,\bn)/a^2)$ in such a way that the
sum admits a finite limit. The relation, valid in two dimensions
\beq
\sqrt{\bg}\, \rR(\bg) \,=\,-\frac12 \Box\ln\det \bg
\eeq
then suggests the introduction of a discretized Laplacian~\footnote{For 
simplicity, we spell out this formula only for $G^{\al\be} = \de^{\al\be}$.},
\beq\label{ven}
Z(a,\bn) \,=\, -  \, \frac12 \ln\left[ 
\frac{ \bg(a(\bn+\bee_0)\big)\bg(a\big(\bn-\bee_0)\big)\bg\big(a(\bn+\bee_1)\big)\bg\big((a(\bn-\bee_1)\big)}{\bg(a\bn)^4}
\right]
\eeq
with lattice unit vectors $\bee_\al$, such that $\lim_{a\ra 0} Z(a,\bn)/a^2 
= \sqrt{\bg}\,\rR(\bg)$.
This procedure therefore amounts to a metric dependent 
`wave-function renormalization' 
  \beq\label{Ren}
\phi(a\bn) \, = \, C Z(a,\bn) \phi_{ren}(a\bn)
\eeq
such that the sum (\ref{f}) is replaced by
\beq
f_a(\phi) \,=\, C a^2 \, \sum_\bn \left(\frac{Z(a,\bn)}{a^2}\right)  \phi_{ren}(a\bn)
\eeq
Therefore we have the renormalized result
\beq\label{Cphi}
\lim_{a\ra  \, 0} \, f_a(\phi) \,=\, 
C \int \rd^2 \si \, \sqrt{\bg}\, \phi_{ren} \rR(\bg)
\eeq
This renormalization prescription is perhaps a bit unusual in view of the fact 
that, in flat space quantum field theory, the wave-function renormalization 
factor depends only on the cutoff, but not on the coordinates. 
However, for a non-trivial background one would expect
the renormalization to also involve the background geometry. 

We emphasize that, up to an overall factor $C$,
this outcome is unique if we demand $(i)$ the final result to be generally
covariant, and $(ii)$ the limit to be such that the $a^2$ factor is properly
taken care of with a finite and well-defined limit. The latter requirement
excludes also higher order derivative contributions such as $\rR^2\,,\,  \rR^3\,, \dots$ .  
Furthermore, by (\ref{f}) there cannot appear any terms with derivatives 
acting on $\phi$ (which would not be accessible to arguments restricted to
constant $\phi$). Luckily, the above renormalization prescription does not affect any other 
terms in  the world-sheet Lagrangian, precisely because the dilaton appears nowhere 
else in the final answer. From our derivation it is clear that if there is any sensible 
result at all for (\ref{W}), it must be proportional to (\ref{phiR}). This is indeed 
all there is to the issue of `renormalization' for this particular sector of the theory!
The overall prefactor  $(4\pi)^{-1}$ in (\ref{phiR}) is then fixed by adjusting the 
proportionality constant in (\ref{Cphi}), and is thus also part of the renormalization prescription (alternatively, its value can be pinned down by arguments
along the lines of \cite{BP,U}). 

As a final comment, we remark that one could also try to apply more
standard heat kernel techniques (see {\em e.g.}  \cite{GMM,Vas}) to the
evaluation of (\ref{W}). More specifically, invoking the Hodge-de Rham decomposition
\beq\label{HdR}
\de A_{\al} \,=\, \pa_\al \om+\eps_{\al\be}\pa^\be \vp
\eeq
we can  change integration variables in (\ref{W}) with the Jacobian
(discarding zero modes)
\beq
\det\, \frac{\de(A_0,A_1)}{\de(\om,\vp)} \,=\,
\det \Box_\bg
\eeq
to recast the exponent of (\ref{W}) in a more familiar form with a 
scalar Laplacian. Inspection of the formulas in section~7.1. of  \cite{Vas} 
then shows that the desired term (\ref{phiR}) does appear, as well as
non-local terms that would be excluded by our arguments above, in
addition to various renormalizations that must be taken into account. However,
apart from the fact that the formulas given there cannot be directly applied to 
the determination of the relevant coefficients for the model at hand, the 
introduction of derivative terms in (\ref{W}) by means of (\ref{HdR}) appears 
rather artificial, in that it obscures the ultralocality of the original expression. 
For this reason we prefer the more direct argument given above.

\section{Outlook}

In summary, we have shown that the dilaton coupling which is missing in the 
double dimensional reduction of the (super-)membrane 
can be accounted for by properly quantizing the membrane, thus completing
the derivation of the world-sheet action for the type I subsector of the full
theory. We have also shown that the construction only works in the critical
dimension, whence non-critical (super-)membrane theories are ruled out.
As expected, the derivation does not extend to the
Ramond-Ramond sector, although our discussion in section~3 does 
clarify why the KK vector field $\cA_\mu$ can only appear in
a gauge invariant combination in the effective target-space Lagrangian,
if it appears at all. The question of how to properly include these  degrees 
of freedom in the world-sheet description of string theory has been under 
discussion for a long time. According to  standard wisdom \cite{Pol} this requires
the extension to open strings and the incorporation of $D$-branes  into the theory. 
By contrast, from the membrane perspective, the world-volume action (\ref{S}) already
includes them in a natural manner from the very outset. It would therefore be 
interesting to relate these two descriptions, but this is a task which we leave to future work.

\vspace{5mm}

\noindent
 {\bf Acknowledgments:} 
We thank Thibault Damour and Jan Derezi{\'n}ski for discussions, and 
Martin Cederwall and  Arkady Tseytlin for comments on a first version of this paper.
K.A.M. was partially supported by the Polish National Science Center 
(grant agreement No UMO-2020/39/B/ST2/01279).
The work of  H.N. has received funding from the European Research 
Council under the  European Union's Horizon 2020 research and 
innovation programme (grant agreement No 740209).


\begin{thebibliography}{99}

\bibitem{BST} 
E.~Bergshoeff, E.~Sezgin and P.K.~Townsend, 
{\it Phys. Lett. B} {\bf 189} (1987) 75.

\bibitem{BST1} 
E.~Bergshoeff, E.~Sezgin and P.K.~Townsend, 
{\it Ann. Phys.} {\bf 185} (1988) 330.

\bibitem{CJS} 
E.~Cremmer, B.~Julia and J.~Scherk, 
{\it Phys. Lett. B} {\bf 76} (1978) 409.

\bibitem{G} 
J.~Goldstone, unpublished.

\bibitem{Hoppe} 
J.~Hoppe,
MIT PhD Thesis, 1982; in: Proc. Int. Workshop on 
{\it Constraint theory and relativistic dynamics}, 
eds.~G.~Longhi and L.~Lusanna (World Scientific, 1987).

\bibitem{dWHN}
B.~de Wit, J.~Hoppe and H.~Nicolai,
{\it Nucl. Phys. B} {\bf 305} (1988) 545.

\bibitem{BFSS}
T.~Banks, W.~Fischler, S.H.~Shenker and L.~Susskind,\\
{\it Phys. Rev. D} {\bf 55} (1997) 5112, {\tt arXiv:hep-th/9610043}.

\bibitem{dWLN} B. de Wit, M. L\"uscher and H. Nicolai, 
Nucl. Phys. {\bf B 320} (1989) 135 

\bibitem{Smilga} A.V. Smilga, 
in Trieste Conference on Supermembranes and Physics in 2 + 1 Dimensions, (1989) 
{\tt 1406.5987 [hep-th]}

\bibitem{HN}
R.~Helling and H.~Nicolai, 
ICTP Spring School 1998 lectures, {\tt  arXiv:hep-th/9809103}.

\bibitem{Duff} 
M.J.~Duff, 
TASI 1996 lectures, {\tt arXiv:hep-th/9611203}.

\bibitem{deWit} 
B.~de Wit, 
1998 Corfu Workshop lectures, {\tt arXiv:hep-th/9902051}.

\bibitem{Banks} 
T.~Banks, 
TASI 1999 lectures, {\tt arXiv:hep-th/9911068}

\bibitem{Taylor} W. Taylor, {\it The M(atrix) model of M-theory}, 
NATO Sci. Ser. C 556 (2000) 91 {\tt hep-th/0002016]}.

\bibitem{FH} J. Fr\"ohlich and J. Hoppe,
Commun. Math. Phys. {\bf 191} (1998) 613 {\tt hep-th/9701119}

\bibitem{MNS} G.~Moore, N. Nekrasov and S. Shatashvili,
Commun. Math. Phys. {\bf 209} (2000) 77

\bibitem{R1}
L.~Boulton, M.~Pilar Garcia del Moral and A.~Restuccia, \\
Nucl. Phys {\bf B671} (2003) 380 {\tt hep-th/0405216}

\bibitem{Hoppe1} J. Hoppe,
Nucl. Phys. {\bf B817} (2009) 155, {\tt 0809.5270 [hep-th]}

\bibitem{R2} 
L.~Boulton, M.~Pilar Garcia del Moral and A.~Restuccia, \\
{\it JHEP} {\bf 05} (2021) 281, {\tt arXiv:2102.00886}.

\bibitem{DNP} 
A.~Dasgupta, H.~Nicolai and J.C.~Plefka, 
{\it JHEP} {\bf 05} (2000) 007, {\tt arXiv:hep-th/0003280}.


\bibitem{LN} O. Lechtenfeld and H. Nicolai, JHEP {\bf 02} (2022) 114, {\tt arXiv:2109.00346}

\bibitem{Duff1} 
M.J.~Duff, P.S.~Howe, T.~Inami and K.S.~Stelle, 
{\it Phys. Lett. B} {\bf 191} (1987) 70.

\bibitem{AKS} A.~Ach\'ucarro, P.~Kapusta and K.S.~Stelle,
Phys. Lett. {\bf B 232} (1989) 302

\bibitem{T} 
P.K.~Townsend, 
{\it Four lectures on M-theory}, 
ICTP summer school 1996 lectures, {\tt arXiv:hep-th/9612121}

\bibitem{Russo} J.G.~Russo,
Nucl. Phys. {\bf B492} (1997) 205

\bibitem{Witten} 
E.~Witten,  
{\it Nucl. Phys. B} {\bf 443} (1995) 85, {\tt arXiv:hep-th/9503124.}

\bibitem{FT}E. S. Fradkin and A. A. Tseytlin, 
Nucl. Phys. {\bf  B261} (1985) 127. [Erratum: Nucl. Phys. {\bf B269} (1986) 745].
    
\bibitem{BP} D.S.~Berman and M.J.~Perry,
Phys. Lett. {\bf B635} (2006) 131,

\bibitem{U} S.~Uehara,
Prog. Theor. Phys. {\bf 124} (2010) 581

\bibitem{MKS} 
U.~Marquard, R.~Kaiser and M.~Scholl, 
Phys. Lett. {\bf B227} (1989) 234.

\bibitem{BdVH} L. Brink, P. Di Vecchia and P. Howe, Phys. Lett. {\bf 65B} (1976) 471

\bibitem{DZ} S. Deser and B. Zumino, Phys. Lett. {\bf 65B} (1976) 369

\bibitem{Ts} A.A.~Tseytlin,
Class. Quant. Grav. {\bf 13} (1996), L81

\bibitem{Callan}   C. G. Callan, Jr., E. J. Martinec, M. J. Perry, and D. Friedan, 
Nucl. Phys. {\bf B262} (1985) 593?609.

\bibitem{Polyakov} A.M.~Polyakov, Phys. Lett. {\bf 103B} (1981) 207

\bibitem{Friedan} D. Z. Friedan, in {\it Recent developments in field theory 
and statistical mechanics} (North-Holland, Amsterdam, 1984)
 
\bibitem{OA} O.~Alvarez, Nucl. Phys. {\bf B216} (1983) 125

\bibitem{MN} G. Moore and P. Nelson, Nucl. Phys. {\bf B266} (1986) 58

\bibitem{FMS} D.Z. Friedan, E.J. Martinec and S. Shenker,
      Nucl. Phys. {\bf B271} (1986) 93

\bibitem{DK} J. Distler and H. Kawai, Nucl. Phys. {\bf B321} (1989) 509

\bibitem{GMM} A.A. Grib, S.G. Mamayev and V.M. Mostepanenko,
{\it Vacuum quantum effects in strong fields},
Friedmann Laboratory Publishing, St. Petersburg (1994)

\bibitem{Vas} D.V.Vassilevich, {\it Heat kernel expansion: user's manual},
  {\tt hep-th/0306138}

\bibitem{Pol} J. Polchinski,
Phys. Rev. Lett. {\bf 75} (1995) 4724;
{\it TASI lectures on D-branes}, {\tt hep-th/9611050}



\end{thebibliography}
\end{document}